\documentclass[12pt,fleqn]{article}
\pdfoutput=1 
\RequirePackage[OT1]{fontenc}
\RequirePackage{amsthm,amsmath}
\RequirePackage[round]{natbib}

\usepackage[font=scriptsize]{caption}
\usepackage{graphicx,psfrag,epsf}
\usepackage{adjustbox}
\usepackage{enumerate}
\usepackage{amsmath}
\usepackage{amsfonts}
\usepackage{amssymb}
\usepackage{amsbsy}
\usepackage[makeroom]{cancel}
\usepackage{bbold}
\usepackage{float}
\usepackage{algorithm}
\usepackage{algpseudocode}
\usepackage{natbib}
\usepackage{url} 
\usepackage[colorlinks=true,
            linkcolor=blue,
            urlcolor=blue,
            citecolor=blue]{hyperref}
\usepackage{etoolbox}
\usepackage{csvsimple}
\usepackage{comment}
\AtBeginEnvironment{tabular}{\small}

\newcommand{\blind}{0}

\begin{document}

\def\spacingset#1{\renewcommand{\baselinestretch}%
{#1}\small\normalsize} \spacingset{1}

\if0\blind
{
  \title{\bf Estimating the marginal likelihood with Integrated nested Laplace approximation (INLA)}
 \author{Aliaksandr Hubin \thanks{
    The authors gratefully acknowledge the \textit{CELS project at the University of Oslo}, \url{http://www.mn.uio.no/math/english/research/groups/cels/index.html}, for giving us the opportunity, inspiration and motivation to write this article.}\hspace{.2cm}\\
    Department of Mathematics, University of Oslo\\
    and \\
    Geir Storvik \\
    Department of Mathematics, University of Oslo}
  \maketitle
} \fi

\if1\blind
{
  \bigskip
  \bigskip
  \bigskip
  \begin{center}
    {\LARGE\bf Estimating the marginal likelihood with INLA}
\end{center}
  \medskip
} \fi

\bigskip
\begin{abstract}

The marginal likelihood is a well established model selection criterion in Bayesian statistics. It also allows to efficiently calculate the marginal posterior model probabilities that can be used for Bayesian model averaging of quantities of interest. For many complex models, including latent modeling approaches, marginal likelihoods are however difficult to compute. One recent promising approach for approximating the marginal likelihood is Integrated Nested Laplace Approximation (INLA), design for models with latent Gaussian structures. In this study we compare the approximations obtained with INLA to some alternative approaches on a number of examples of different complexity. In particular we address a simple linear latent model, a Bayesian linear regression model, logistic Bayesian regression models with probit and logit links, and a Poisson longitudinal generalized linear mixed model. 

\end{abstract}

\noindent%
\vfill

\spacingset{1.2} 

\section{Introduction}

Marginal likelihoods have been commonly accepted to be an extremely important quantity within Bayesian statistics. For data $\mathbf{y}$ and model  $\mathcal{M}$, which includes some unknown parameters $\theta$, the marginal likelihood is given by
\begin{equation}
p(\mathbf{y}|\mathcal{M})=\int_{\Omega_\theta}p(\mathbf{y}|\mathcal{M},\theta)p(\theta|\mathcal{M})d\theta\label{mlikdef}
\end{equation}
where $p(\theta|\mathcal{M})$ is the prior for $\theta$  under model $\mathcal{M}$ while $p(\mathbf{y}|\mathcal{M},\theta)$ is the likelihood function conditional on $\theta$. 
Consider first the problem of comparing models $\mathcal{M}_i$ and $\mathcal{M}_j$ through the ratio between their posterior probabilities:
\begin{equation}
\frac{p(\mathcal{M}_i|\mathbf{y})}{p(\mathcal{M}_j|\mathbf{y})} = \frac{p(\mathbf{y}|\mathcal{M}_i)}{p(\mathbf{y}|\mathcal{M}_j)}\times \frac{p(\mathcal{M}_i)}{p(\mathcal{M}_j)}.\label{BF}
\end{equation}
The first term of the right hand side is the Bayes Factor~\citep{kass1995bayes}.
In this way one usually performs Bayesian model selection with respect to the posterior marginal model model probabilities without the need to calculate them explicitly. However if we are interested in Bayesian model averaging and marginalizing some quantity $\Delta$ over the given set of models $\Omega_{\mathcal{M}}$ we are calculating the posterior marginal distribution, which in our notation becomes:
\begin{equation}\label{posterior_quantile}
p(\Delta|\mathbf{y}) =  \sum_{\mathcal{M} \in \Omega_{\mathcal{M}}}{p(\Delta|\mathcal{M},\mathbf{y})p(\mathcal{M}|\mathbf{y})}.
\end{equation}
Here $p(\mathcal{M}|\mathbf{y})$ is the posterior marginal model probability for model $\mathcal{M}$ that can be calculated with respect to Bayes theorem as:
\begin{equation}\label{PMP}
p(\mathcal{M}|\mathbf{y}) =  \frac{{p(\mathbf{y}|\mathcal{M})p(\mathcal{M})}}{\sum_{\boldsymbol{\mathcal{M}}' \in\Omega_{\mathcal{M}}}{p(\mathbf{y}| \mathcal{M}')p(\mathcal{M}')}},
\end{equation}
Thus one requires marginal likelihoods $p(\mathbf{y}|\mathcal{M})$ in \eqref{BF}, \eqref{posterior_quantile} and \eqref{PMP}.  Metropolis-Hastings algorithms searching through models within a Monte Carlo setting~\citep[e.g.][]{Hubin2016} requires acceptance ratios of the form
\begin{equation}
r_m({\mathcal{M}},\mathcal{M}^*)=\min\left\lbrace1,\frac{p(\mathbf{y}|\mathcal{M}^*)p(\mathcal{M}^*)q({\mathcal{M}}|\mathcal{M}^*)}{p(\mathbf{y}|\mathcal{M})p(\mathcal{M})q(\mathcal{M}^*|{\mathcal{M}})}\right\rbrace\label{balance}
\end{equation}
also involving the marginal likelihoods. All these examples show the fundamental importance of being able to calculate marginal likelihoods $p(\mathbf{y}|\mathcal{M})$  in Bayesian statistics.

Unfortunately for most of the models that include both unknown parameters $\theta$ and some latent variables $\boldsymbol{\eta}$  analytical calculation of $p(\mathbf{y}|\mathcal{M})$ is impossible.  In such situations one must use approximative methods that hopefully are accurate enough to neglect the approximation errors involved. Different approximative approaches have been mentioned in various settings of Bayesian variable selection and Bayesian model averaging. Laplace's method \citep{tierney1986accurate} has been widely used, but it is based on rather strong assumptions. The Harmonic mean estimator~\citep{newton1994approximate} is an easy to implement MCMC based method, but it can give high variability in the estimates.
Chib's method~\citep{chib1995marginal}, and its extension~\citep{chib2001marginal}, are also MCMC based approaches that have gained increasing popularity. They can be very accurate provided enough MCMC iterations are performed, but need to be adopted to each application and the specific algorithm used. Approximate Bayesian Computation~\citep[ABC,][]{marin2012approximate} has also been considered in this context, being much faster than MCMC alternatives, but also giving cruder approximations.
Variational methods~\citep{jordan1999introduction} provide lower bounds for the marginal likelihoods and have been used for model selection in e.g. mixture models~\citep{mcgrory2007variational}.
Integrated nested Laplace approximation~\citep[INLA,][]{rue2009eINLA} provides estimates of marginal likelihoods within the class of latent Gaussian models and has become extremely popular. The reason for it is that Bayesian inference within INLA is extremely fast and remains at the same time reasonably precise. 

\citet{Friel2012} perform comparison of some of the mentioned approaches including Laplace approximations, harmonic mean approximations, Chib's method and other. However to our awareness there were no studies comparing the approximations of the marginal likelihood obtained by INLA with other popular methods mentioned in this paragraph. Hence the main goal of this article is to explore the precision of  INLA in comparison with the mentioned above alternatives. INLA approximates marginal likelihoods by
\begin{equation}
p(\mathbf{y}|\mathcal{M})\approx \int_{\Omega_\theta} \left. \frac{p(\mathbf{y},\theta,\boldsymbol{\eta}|\mathcal{M})}{\tilde{\pi}_G(\boldsymbol{\eta}|\mathbf{y},\theta,\mathcal{M})} \right \rvert_{\boldsymbol{\eta} = \boldsymbol{\eta}^*(\theta|\mathcal{M})} d\theta,
\end{equation}
where $\boldsymbol{\eta}^*(\theta|\mathcal{M}) $ is some chosen value of $\boldsymbol{\eta}$, typically the posterior mode, while $\tilde{\pi}_G(\boldsymbol{\eta}|\mathbf{y},\theta,\mathcal{M})$ is a Gaussian approximation to $\pi(\boldsymbol{\eta}|\mathbf{y},\theta,\mathcal{M})$. Within the INLA framework both random effects and regression parameters are treated as latent variables, making the dimension of the hyperparmeters $\theta$ typically low. The integration of $\theta$ over the support ${\Omega_\theta}$ can be performed by an empirical Bayes (EB) approximation or using numerical integration based on a central composite design (CCD) or a grid~\citep[see][for details]{rue2009eINLA}. 

In the following sections we will evaluate the performance of INLA through a number of examples of different complexities, beginning with a simple linear latent model and ending up with a Poisson longitudinal generalized linear mixed model.

\section{INLA versus truth and the harmonic mean}
To begin with we address an extremely simple example suggested by~\citet{NealBlog}, in which we consider the following model $\mathcal{M}$:
\begin{equation}
\begin{split}
Y|\eta,\mathcal{M}\sim& N(\eta,\sigma_1^2);\\
\eta|\mathcal{M}\sim& N(0,\sigma_0^2).
\end{split}
\end{equation}
Then obviously the marginal likelihood is available analytically as
\begin{eqnarray*}
&Y|\mathcal{M} \sim N(0,\sigma_0^2+\sigma_1^2),
\end{eqnarray*}
and we have a benchmark to compare approximations to. The harmonic mean estimator~\citep{raftery2006estimating} is given by
\begin{equation*}
p(y|\mathcal{M})\approx \frac{n}{\sum_{i=1}^{n}\frac{1}{p(y|\eta_i,\mathcal{M})}}
\end{equation*}
where $\eta_i\sim p(\eta|y,\mathcal{M})$. 
This estimator is consistent, however often requires too many iterations to converge. We performed the experiments with $\sigma_1 = 1$ and $\sigma_0$ being either 1000, 10 or 0.1. The harmonic mean is obtained based on $n = 10^7$ simulations and 5 runs of the harmonic mean procedure are performed for each scenario. For INLA we used the default tuning parameters from the package (in this simple example different settings all give equivalent results).
\begin{table}[h]
\begin{center}
\begin{tabular}{|c|c|c|c|c|rrrrr|}
\hline
$\sigma_0$&$\sigma_1$&$D$&Exact&INLA&\multicolumn{5}{c|}{H.mean}\\\hline
1000&1&2&-7.8267&-7.8267&-2.4442&-2.4302& -2.5365&-2.4154&-2.4365\\\hline
10&1&2&-3.2463&-3.2463&-2.3130&-2.3248&-2.5177&-2.4193&-2.3960 \\\hline 
0.1&1&2&-2.9041&-2.9041&-2.9041&-2.9041&-2.9042& -2.9041&-2.9042\\\hline
\end{tabular}
\end{center}
\caption{Comparison of INLA, harmonic mean and exact marginal likelihood}\label{mlikcomp1}
\end{table}
As one can see from Table \ref{mlikcomp1} INLA gives extremely precise results even for a huge variance of the latent variable, whilst the harmonic mean can often become extremely crude even for $10^7$ iterations.  Due to the bad performance of the harmonic mean~\citep[see also][]{NealBlog} this method will not be considered further.

\section{INLA versus Chib's method in Gaussian Bayesian regression}
In the second example we address INLA versus Chib's method \citep{chib1995marginal} for the US crime data~\citep{Vandaele1978} based on the following model $\mathcal{M}$:
\begin{equation}\label{normgammf}
\begin{split}
 Y_t|\mu_t,\mathcal{M} \stackrel{iid}{\sim}&  N(\mu_t,\sigma^2);\\
 \mu_t|\mathcal{M}=& \beta_0 + \sum_{i=1}^{p_{\mathcal{M}}} \beta_{i}x^{\mathcal{M}}_{ti};\\
  \frac{1}{\sigma^2}|\mathcal{M} \sim& \text{ Gamma}(\alpha_\sigma,\beta_\sigma);\\
  \beta_i|\mathcal{M} \sim& N(\mu_\beta,\sigma_{\beta}^2),
  \end{split}
\end{equation}
where $t=1,...,47$ and $i=0,...,p_{\mathcal{M}}$.
We also addressed two different models, $\mathcal{M}_1$ and $\mathcal{M}_2$, induced by different sets of the explanatory variables with cardinalities $p_{\mathcal{M}_1}=8$ and $p_{\mathcal{M}_2}=11$ respectively.

\begin{figure}[h]
\centering
\includegraphics[width=0.45\linewidth]{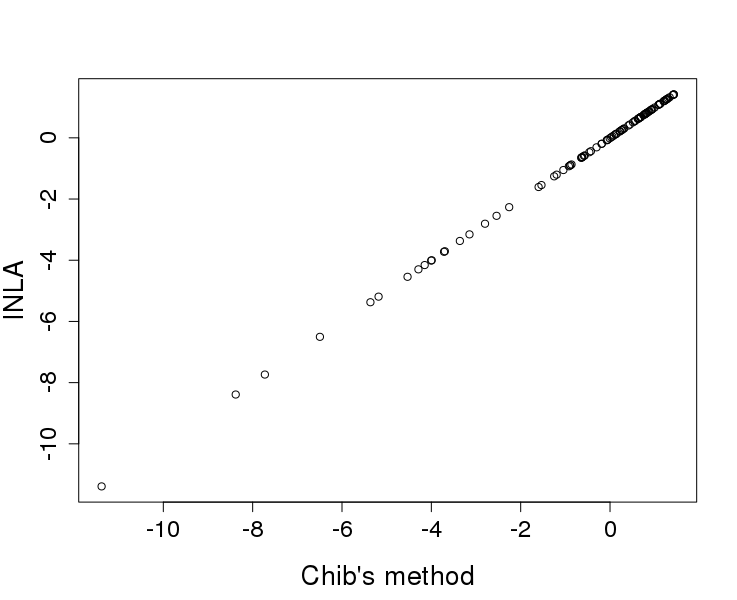}
\includegraphics[width=0.45\linewidth]{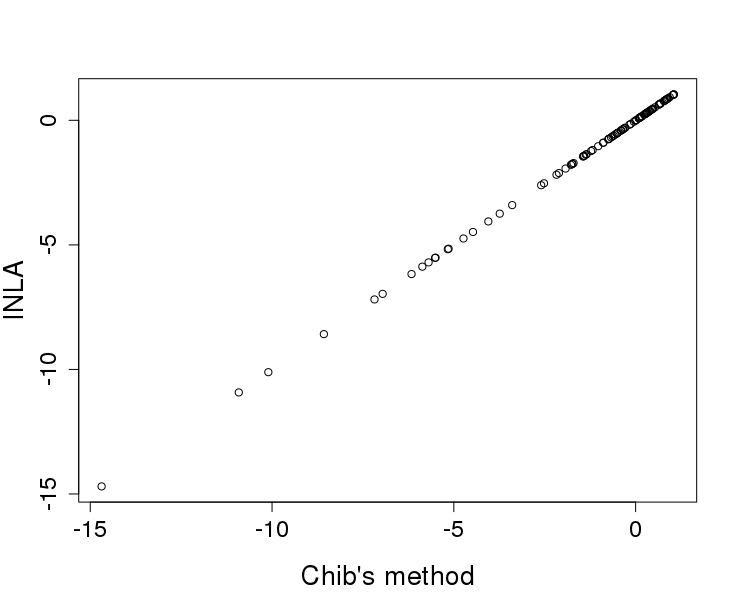}
\caption{Chib's-INLA plots of estimated marginal log-likelihoods obtained by Chib's method ($x$-axis) and INLA ($y$-axis) for 100 different values of $\sigma_{\beta}^{-2}$. The left plot corresponds to model $\mathcal{M}_1$ while the right plot corresponds to model $\mathcal{M}_2$\label{chinla0}}
\end{figure}

In model~\eqref{normgammf} the hyperparameters were specified to $\mu_\beta = 0,\alpha_\sigma = 1$ and $\beta_\sigma = 1$.
Different precisions $\sigma_{\beta}^{-2}$ in the range $[0,100]$ were tried out in order to explore the properties of the different methods with respect to prior settings. Figure~\ref{chinla0} shows the estimated marginal log-likelihoods for Chib's method ($x$-axis) and INLA ($y$-axis) for model  $\mathcal{M}_1$ (left) and $\mathcal{M}_2$ (right). Essentially, the two methods give equal marginal likelihoods in each scenario.
Table~\ref{mlikcomp2} shows more details for a few chosen values of the standard deviation $\sigma_{\beta}$. The means of the 5 replications of Chib's method all agree with INLA up to the second decimal.
\begin{table}[h]
\begin{center}
\begin{tabular}{|l|c|c|c|rrrrr|}
\hline
$\mathcal{M}$&$\mu_\beta$&$\sigma_\beta$&INLA&\multicolumn{5}{c|}{Chib's method}\\\hline
$\mathcal{M}_1$&0&1000&-73.2173&-73.2091&-73.2098&-73.2090&-73.2088&-73.2094\\\hline
$\mathcal{M}_1$&0&10&-31.7814&-31.7727&-31.7732&-31.7732&-31.7725&-31.7733\\\hline
$\mathcal{M}_1$&0&0.1&1.4288&1.4379&1.4380&1.4383&1.4378&1.4376\\\hline
$\mathcal{M}_2$&0&1000&-96.6449&-96.6372&-96.6368&-96.6370&-96.6373&-96.6370\\\hline
$\mathcal{M}_2$&0&10&-41.4064&-41.3989&-41.3987&-41.3991&-41.3995&-41.3996\\\hline
$\mathcal{M}_2$&0&0.1&1.0536&1.0625&1.0629&1.0628&1.0626&1.0625\\\hline
\end{tabular}
\end{center}
\caption{Comparison of INLA and Chib's method for marginal log likelihood.}\label{mlikcomp2}
\end{table}

\begin{figure}[h]
\centering
\includegraphics[width=\linewidth]{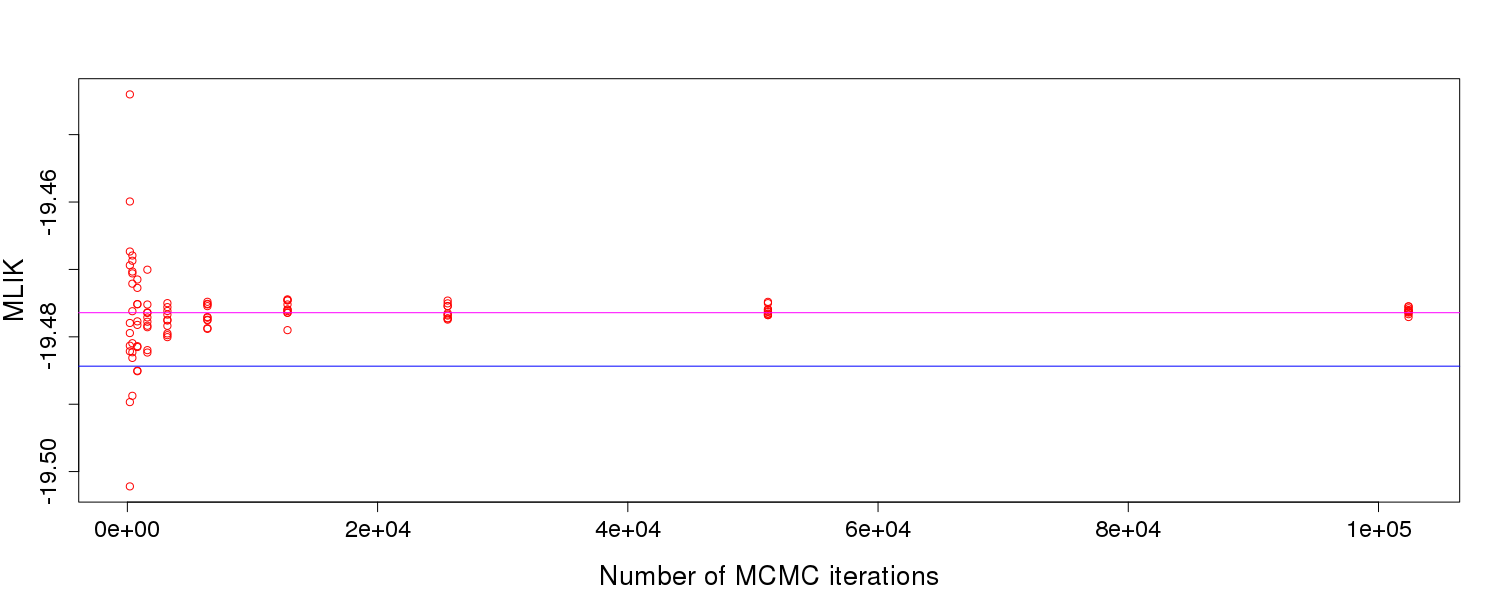}
\caption{Variability  of Chib's method as a function of number of MCMC iterations for the simple linear Gaussian example. Horizontal lines correspond to INLA estimates based on default settings (dark blue) and adjusted settings (purple).}\label{figadj}
\end{figure}

Going a bit more into details, Figure~\ref{figadj} shows the performance of Chib's method as a function of the number of iterations. 
The red circles in this graph represent 10 runs of Chib's method for several choices of the number of iterations of the algorithm changing from 200 to 102400.
The horizontal solid line shows the INLA estimate with default settings. In this case, we used a precision on the regression parameters equal to $\sigma^{-2}_\beta = 0.2$, while in order to obtain some difference between Chib's method and INLA we changed the mean to $\mu_\beta = 1$. We only considered model $\mathcal{M}_1$ in this case.
Although the differences are still small, this illustrates that INLA can be a bit off the true value. The reason for this deviance is due to the default choice of values for the tuning parameters in INLA. After tuning the step of numerical integration $\delta_z$  defining the grid as well as the convergence criterion of the differences of the log densities $\pi_z$~\citep{rue2009eINLA} one can make the difference between INLA and Chib's method arbitrary small for this example. This can be clearly seen in Figure \ref{figadj}, where we depict the default INLA results (dark blue line) and the tuned INLA results (purple line). 
From Figure~\ref{figadj} one can also see that it might take quite a while for Chib's method to converge, whilst INLA gives stable results for the fixed values of the tuning parameters. The total computational time for INLA corresponds to about 50\,000 iterations with Chib's method for this model. Whilst 819200 iterations of Chib's method would require at least 15 times more time than INLA on the same machine \footnote{Intel(R) Core(TM) i5-6500 CPU @ 3.20GHz with 16 GB RAM was used for all of the computations}.

The main conclusion that can be drawn from this example is that INLA approximations of marginal likelihoods can indeed be trusted for this model,  giving yet another evidence in the support of INLA methodology in general.

\section{INLA versus Chib's method for logistic Bayesian regression with a probit link}
\begin{figure}[h]
\centering
\includegraphics[width=0.45\linewidth]{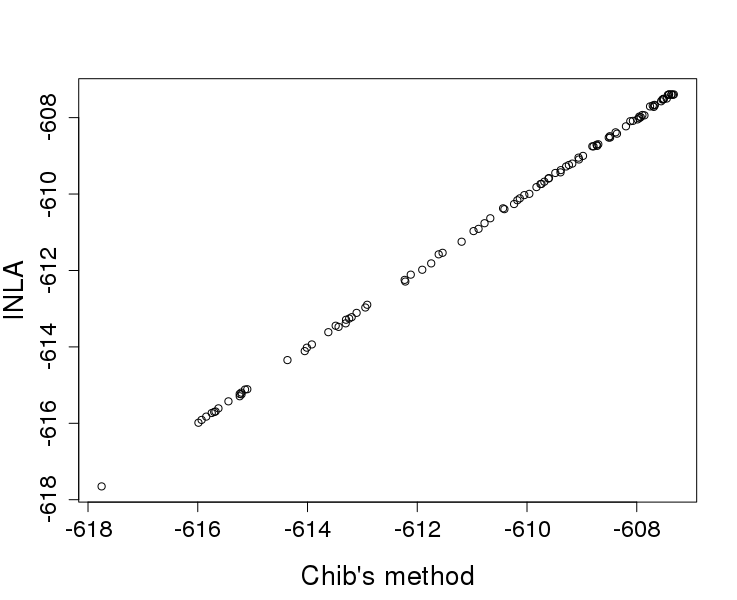}
\includegraphics[width=0.45\linewidth]{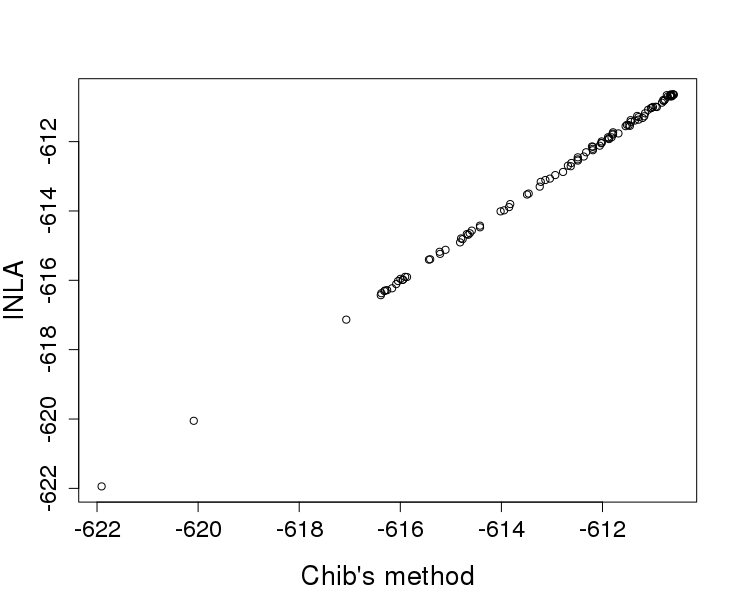}
\caption{Comparisons of marginal likelihood estimates obtained by Chib's method ($x$-axis) and INLA ($y$-axis) for 100 different values of the precision parameter $\sigma_{\beta}^{-2}$  under 2 models with different number of covariates.}\label{ldiffml}
\end{figure}

In the third example we will continue comparing INLA with the Chib's method \citep{chib1995marginal} for approximating the marginal likelihood in logistic regression with a probit link model $\mathcal{M}$.
 The data set addressed is the simulated Bernoulli data introduced by \citet{Hubin2016}. The model is given by
\begin{align}
\begin{split}
 Y_t|p_t,\mathcal{M} \stackrel{iid}{\sim}&  \text{ Bernoulli}(\Phi
 (\eta_t));\\
 \eta_t|\mathcal{M}  =& \beta_0 + \sum_{i=1}^{p_{\mathcal{M} }} \beta_{i}x^{\mathcal{M}}_{ti};\\
 \beta_i|\mathcal{M}  \sim& N(\mu_\beta,\sigma_{\beta}^2),
 \end{split}
\end{align}
where $t=1,...,2000$ and $i=0,...,p_{\mathcal{M}}$.
We addressed two different sets of explanatory variables with different cardinalities of 11 for model $\mathcal{M}_1$ and 13 for model $\mathcal{M}_2$.
\begin{table}[h]
\begin{center}
\begin{tabular}{|l|c|c|c|rrrrr|}
\hline
$\mathcal{M}$&$\mu_\beta$&$\sigma_\beta$&INLA&\multicolumn{5}{c|}{Chib's method}\\\hline
$\mathcal{M}_1$&0&1000&-688.3192&-688.2463&-688.3260&-688.3117&-688.2613&-688.2990\\\hline
$\mathcal{M}_1$&0&10&-633.0902&-633.1584&-633.0612&-633.0335&-633.1094&-633.0780\\\hline
$\mathcal{M}_1$&0&0.1&-669.7590&-669.7646&-669.7666&-669.7610&-669.7465&-669.7528\\\hline
$\mathcal{M}_2$&0&1000&-704.2266&-704.2154&-704.2138&-704.1463&-704.2526&-704.2303\\\hline
$\mathcal{M}_2$&0&10&-639.8051&-639.7932&-639.8349&-639.8022&-639.7675&-639.8278\\\hline
$\mathcal{M}_2$&0&0.1&-649.7803&-649.7360&-649.7604&-649.7893&-649.7532&-649.7806\\\hline
\end{tabular}
\end{center}
\caption{Comparison of INLA and Chib's method for logistic Bayesian regression with a probit link}\label{mlikcomp3}
\end{table}
We used $\mu_\beta = 0$ while the precisions for the
regression parameters were varied between 0 and 10 in Figure~\ref{ldiffml} and chosen as $10^{-6}, 10^{-2}$ and $10^2$ in Table~\ref{mlikcomp3}.
Figure~\ref{ldiffml} shows that INLA and Chib's method give reasonably similar results for both models.
The total time for running INLA within these examples is at most  2 seconds, corresponding to approximately 12000 MCMC iterations in Chib's method.
100\,000 MCMC iterations that were used to produce the obtained results in Table \ref{mlikcomp3} required at least  25 seconds per replication on the same machine.
\begin{figure}[h]
\centering
\includegraphics[width=\linewidth]{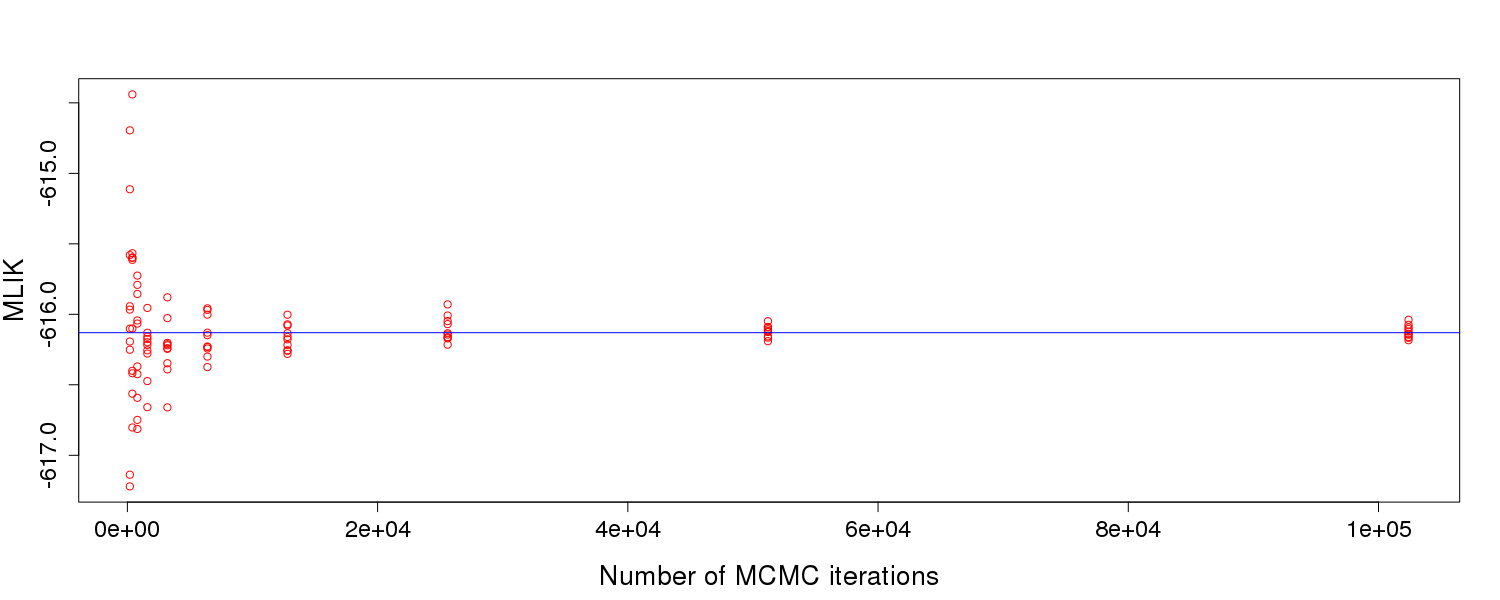}
\caption{Variability  of Chib's method as a function of number of MCMC iterations for logistic Bayesian regression with a probit link. The horizontal line corresponds to the INLA estimate based on default settings.}\label{figadj2}
\end{figure}
\clearpage

\section{INLA versus other methods for logistic Bayesian regression with a logit link}
In the fourth example we will continue comparing marginal likelihoods obtained by INLA with such methods as Laplace approximations, Chib and Jeliazkov's method, Laplace MAP approximations, harmonic mean method, power posteriors, annealed importance sampling and nested sampling. The model $\mathcal{M}$ is the Bayesian logistic regression model addressed by \citet{Friel2012}, which is given by
\begin{align}
\begin{split}
 Y_t|p_t,\mathcal{M} \stackrel{iid}{\sim}&  \text{ Bernoulli}(\text{logit}^{-1}(\eta_t));\\
 \eta_t|\mathcal{M}  =&\beta_0 + \sum_{i=1}^{p_{\mathcal{M} }} \beta_{i}x^{\mathcal{M}}_{ti};\\
 \beta_i|\mathcal{M}  \sim& N(\mu_\beta,\sigma_{\beta}^2),
 \end{split}
\end{align}
where $t=1,...,532$ and $i=0,...,p_{\mathcal{M}}$. The data set addressed is the Pima Indians data, which consist of some diabetes records for $532$ Pima Indian women of different ages. For $\mathcal{M}_1$ we have addressed such predictors as the number of pregnancies, plasma glucose concentration, body mass index and diabetes pedigree function and for $\mathcal{M}_2$ we additionally consider the age covariate.  All of the covariates for both of the models have been standardized before the analysis. Then the analysis was performed for $\sigma_{\beta}^2=100$ and  $\sigma_{\beta}^2=1$ correspondingly. The prior value of $\mu_\beta$ for both of the cases was chosen to be equal to 1. Table \ref{mlikcomp4} contains the results obtained by all of the methods. Notice that all of the calculations apart from the INLA based ones are reported in  \citet{Friel2012}. \citet{Friel2012} claim that the relevant measures were taken to make the implementation of each method as fair as possible. In their runs each Monte Carlo method used the equivalent of 200\,000 samples. In particular, the power posteriors used 20\,000 samples at each of the 10 steps. The annealed importance sampling a cooling scheme with 100 temperatures and 2\,000 samples generated per temperature. Nested sampling was allowed to use 2\,000 samples and was terminated when the contribution to the current value of marginal likelihood was smaller than $10^{-8}$ times the current value. Notice that the default tuning parameters were applied for the INLA calculations. Except for the Harmonic mean, all methods gave comparable results. The INLA method only needed a computational time comparable to Laplace approximations, which is much faster than the competing approaches~\citep{Friel2012}. Reasonably good performance of the ordinary Laplace approximation in this case can be explained by having no latent variables in the model.

\begin{table}[h]
\begin{center}
\begin{tabular}{lcccc}
\hline
Method&$\mathcal{M}_1$&$\mathcal{M}_2$&$\mathcal{M}_1$&$\mathcal{M}_2$\\\hline
INLA&-257.25&-259.89&-247.32&-247.59\\
Laplace approximation&-257.26&-259.89&-247.33&-247.59\\
Chib and Jeliazkov's method&-257.23&-259.84&-247.31&-247.58\\
Laplace approximation MAP&-257.28&-259.90&-247.33&-247.62\\
Harmonic mean estimator&-279.47&-284.78&-259.84&-260.55\\
Power posteriors&-257.98&-260.59&-247.57&-247.84\\
Annealed importance sampling&-257.87&-260.43&-247.30&-247.59\\
Nested sampling&-258.82&-261.38&-246.82&-246.97
\\\hline
 $\sigma_{\beta}^2 $ value&100&100&1&1\\\hline
\end{tabular}
\end{center}
\caption{Comparison of INLA and other method for a logistic Bayesian regression with a logit link.}\label{mlikcomp4}
\end{table}

\section{INLA versus Chib and Jeliazkov's method for computation of marginal likelihoods in a Poisson with a mixed effect model}
As models become more sophisticated we have less methodologies that can be used for approximating the marginal likelihood. In the context of generalized linear mixed models two alternatives will be considered, the INLA approach \citep{rue2009eINLA} and the Chib and Jeliazkov's approach~\citep{chib2001marginal}. 

 This model is concerned with seizure counts $Y_{jt}$ for 59 epileptics measured first over an 8-week baseline period $t=0$ and then over 4 subsequent
2-week periods $t =1,...,4$. After the baseline period
each patient is randomly assigned to either receive a specific drug or a placebo. Following previous analyses of these data, we removed observation 49, considered to be an outlier because of the unusual seizure counts. We assume the data to be Poisson distributed and model both fixed and random effects of based on some covariates. The model $\mathcal{M}$,  originally defined in~\citet{diggle1994analysis}, is given by
\begin{align}
\begin{split}
 Y_{jt}|\lambda_{jt},\mathcal{M} \sim&  \text{ Poisson}(\exp(\eta_{jt}));\\
 \eta_{jt}|\mathcal{M} =& \log(\tau_{jt})+\beta_0 + \beta_1x_{jt1} + \beta_1x_{jt2} + \beta_3x_{jt1}x_{jt2} +
   b_{j0} + b_{j1}x_{jt1};\\
 \beta_i|\mathcal{M} \sim& N(0,100),\quad\quad i=0,...,3;\\
 \bold{b_j}|\mathbf{D},\mathcal{M} \sim& N_2(0,\mathbf{D});\\
 \mathbf{D}^{-1}|\mathcal{M} \sim& \text{ Wishart}_2(4,I_2),
 \end{split}
\end{align}
for $j=1,...,58,t=1,...,4$. Here $x_{jt1} \in \{0,1\}$ is an indicator variable of period (0 if baseline and 1 otherwise), $x_{jt2} \in \{0,1\}$ is an indicator for treatment status, $\tau_{it}$ is the offset that is equal to 8 in the baseline period and 2
otherwise, and $\bold{b_j}$ are latent random effects. In~\citet{chib2001marginal} an estimate of the marginal log-likelihood was reported to be -915.49, while also an alternative estimate equal to -915.23 based on a kernel density approach by~\citet{Chib1998} was given. INLA gave a value of -915.61 in this case, again demonstrating its accuracy. The computational time for the INLA computation was in this case on average 1.85 seconds.

%

\section{Conclusions}
The marginal likelihood is a fundamental quantity in the Bayesian statistics, which is extensively adopted for Bayesian model selection and averaging in various settings. In this study we have compared the INLA methodology to some other approaches for approximate calculation of the marginal likelihood. In all of the addressed examples disregarding complexity of the latter INLA gave reliable estimates. In all cases, default settings of the INLA procedure gave reasonable accurate results. If extremely high accuracy is needed we recommend that before performing Bayesian model selection and averaging in a particular model space $\Omega_\mathcal{M}$ based on marginal likelihoods produced by INLA, the produced estimates should be carefully studied and the tuning parameters adjusted, if required. Experimenting with different settings will also give an indication on whether more accuracy is needed. 

\bigskip
\begin{center}
{\large\bf SUPPLEMENTARY MATERIAL}
\end{center}
\textbf{Data and code:} Data (simulated and real) and \textit{R} scripts for calculating marginal likelihoods under various scenarios are available online at \url{https://goo.gl/0Wsqgp}. \\

\footnotesize

\let\oldbibliography\thebibliography
\renewcommand{\thebibliography}[1]{\oldbibliography{#1}
\setlength{\itemsep}{0pt}} 

\bigskip
\begin{center}
{\large\bf ACKNOWLEDGMENTS}
\end{center}

We would like to thank CELS project at the University of Oslo for giving us the opportunity, inspiration and motivation to write this article.


%
%


\bibliography{ref}
\bibliographystyle{abbrvnat}

\end{document}